\documentclass[aip,reprint]{revtex4-1}

\draft % marks overfull lines with a black rule on the right
\usepackage[latin1]{inputenc}
\usepackage[pdftex]{graphicx}
\usepackage{soul,color}
\usepackage{textcomp}
\usepackage{hyperref}
\usepackage{amssymb}
\usepackage{amsmath}
\usepackage{ifthen} % provides \ifthenelse test  
\usepackage{xifthen} % provides \isempty test

\newcommand{\fig}[2][]{%
\ifthenelse{\isempty{#1}}
{Fig.~\ref{#2}}% if no subfigure is given
{Fig.~\ref{#2}(#1)}% else
}

\begin{document}

%\preprint{}

\title{Strong 4-mode coupling of nanomechanical string resonators} %Title of paper



\author{Katrin Gajo}
\author{Simon Sch\"uz}
\author{Eva M. Weig}
\email[]{eva.weig@uni-konstanz.de}

%\thanks{}
\affiliation{University of Konstanz, Department of Physics, 78457 Konstanz, Germany}


\date{\today}

\begin{abstract}
We investigate mechanical mode coupling between the four fundamental flexural modes of two doubly-clamped, high-Q silicon-nitride nanomechanical string resonators. Strong mechanical coupling between the strings is induced by the strain mediated via a shared clamping point, engineered to increase the exchange of oscillatory energy. One of the resonators is controlled dielectrically, which results in strong coupling between its out-of-plane and in-plane flexural modes. We show both, inter-string out-of-plane-in-plane and 3-mode resonance of the four coupled fundamental vibrational modes of a resonator pair, giving rise to a simple and a multimode avoided crossing, respectively.

\end{abstract}


\maketitle
\begin{figure}[t]
\includegraphics{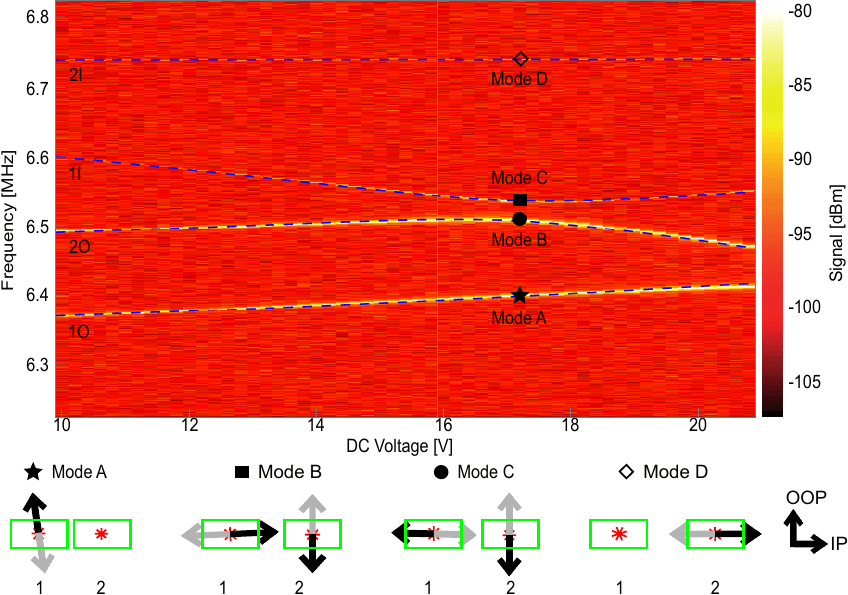}
\caption{\label{2}(color online) Two-mode resonance: DC voltage sweep of resonator 1 shows avoided crossing between OOP mode of resonator 2 (mode 2O) and IP mode of resonator 1 (mode 1I), leading to the hybridized normal modes B and C, while the OOP mode of resonator 1 (1O) and the IP mode of resonator (2I) remain largely unaffected (mode A and D, respectively). Color scale shows logarithmic signal power. The fit of the analytic model (blue dashed lines) yields a coupling strength of $g_{2O,1I}/2\pi$=\,29.7\,kHz. The black arrows show the respective mode polarization of modes A, B, C and D at a DC voltage of 17.2\,V  ($\updownarrow$ : OOP, $\leftrightarrow$ : IP), grey arrows show the inverse to illustrate the vibration of the strings.}
\end{figure}

 %\maketitle must follow title, authors, abstract and \pacs
In recent years strain coupling has been explored as a versatile means to couple the flexural modes of a nanomechanical resonator to other degrees of freedom\cite{Ovartchaiyapong_Jayich,Yeo_Arcizet_2014,Lee_Jayich}, as well as to other mechanical modes\cite{Venstra_vanderZant_2010,Westra_Venstra_2010,Lulla_Owers-Bradley_2012,Eichler_Bachtold_2012}. While strain coupling is usually relatively weak\cite{Venstra_vanderZant_2010,Westra_Venstra_2010,Lulla_Owers-Bradley_2012}, the strong coupling regime of two nanomechanical modes has been achieved taking advantage of internal resonance\cite{Eichler_Bachtold_2012}. Alternatively, strong coupling of strain coupled modes has been accomplished by additionally establishing a dynamical, parametric coupling\cite{Venstra_vanderZant_2011,Mahboob_Yamaguchi_2013,Mathew_Deshmukh_2015,Okamoto_Yamaguchi_2013,Mahboob_Yamaguchi_2014,Okamoto_Kippenberg}, even though we note that this is not in the focus of the present manuscript.
Strongly coupled modes offer interesting prospects for sensing applications\cite{Lepinay_Arcizet_2016,Rossi_Poggio_2017}. Furthermore they constitute a mechanical two-mode system\cite{Faust_2012,Okamoto_Yamaguchi_2013,Faust_Weig_2013} which allows to address and explore quantum-classical analogies. In addition, the nonlinear response of strongly coupled modes allows to study injection locking phenomena\cite{Gil-Santos_Favero_2017,Mahboob_Yamaguchi_2014,Seitner_Weig_2017} and chaos\cite{Karabalin_Roukes_2011}. While many of these effects can be studied using the strongly coupled modes of a single resonator\citep{Faust_2012,Faust_Weig_2013}, the strong coupling of different resonators\citep{Okamoto_Yamaguchi_2013} may be advantageous for addressing the individual subsystem. In particular, strain-mediated coupling schemes, which have previously been mainly employed to realize modal coupling within a single resonator\cite{Eichler_Bachtold_2012,Matheny_Roukes_2013,Venstra_vanderZant_2010,Westra_Venstra_2010,Lulla_Owers-Bradley_2012} can be adapted to enable inter-resonator coupling\cite{Karabalin_Roukes_2011,Huebl}. Even more the possibility to couple adjacent resonators\cite{Okamoto_Yamaguchi_2013,Matheny_Roukes_2014,Agrawal_Seshia_2014} allows for scaling towards resonator arrays\cite{Buks_Roukes_2002,Huang_Du_2016,Okamoto_Kippenberg,Kambali_2015} which entails the prospect to achieve multimode coupling architectures.
\begin{figure}[t]
\includegraphics{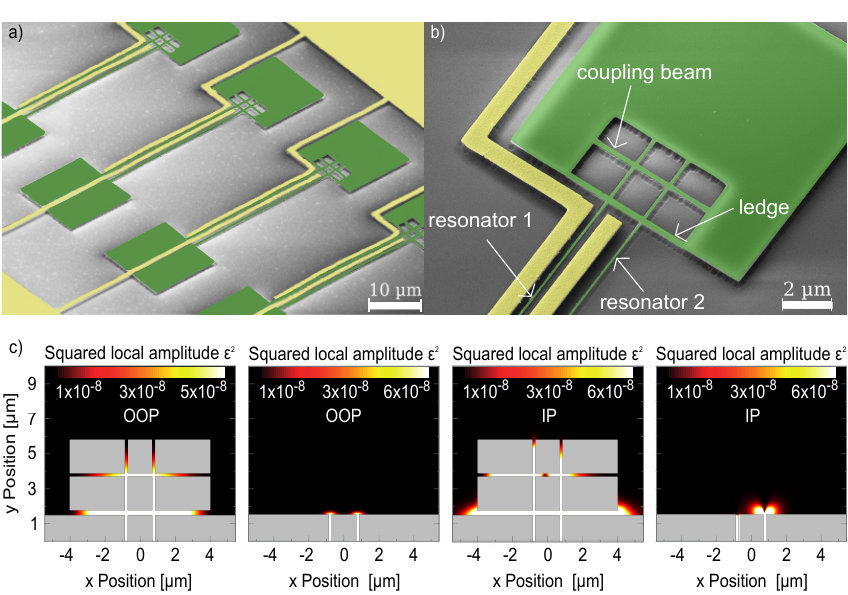}
\caption{\label{1}(color online). a) False color SEM micrograph of an array of pairs of freely suspended doubly clamped silicon nitride string resonators (green) with gold electrodes (yellow) for dielectric actuation and frequency tuning. b) Shared clamping region of a pair of resonators showing the coupling window including the coupling beam and the window ledge. The undercut of the device is 300\,nm. All parts of the window structure are freely suspended. c) Finite element simulation of the squared normalized local amplitude $\epsilon^2$ (color scale) of the vibrational strain field in the clamping region for OOP and IP fundamental modes with coupling window (left) shows enhanced strain field extension compared to the layout without a window (right), respectively.}
\end{figure}\\
Here we present strong intrinsic coupling of the four fundamental vibrational modes of a pair of  two parallel, doubly clamped, high-stress ($\sim$1.46\,GPa) silicon-nitride string resonators sharing one clamping point. One of the two resonators is flanked by a pair of adjacent gold electrodes for dielectric actuation\cite{2009_Unterreitheimer_Kotthaus} and tuning\cite{Rieger_Weig_2012} of both in-plane (IP) and out-of-plane (OOP) modes of the respective resonator, while the eigenmodes of the second resonator remain largely unaffected. The shared clamping point is engineered to support the exchange of vibrational energy from the driven resonator (resonator 1) to the non-driven resonator (resonator 2). This is accomplished by tailoring the mechanical impedance mismatch between resonator and clamping point\cite{Rieger_Weig_2014} to enable the strain fields of both to overlap in the clamping region.\\
Using DC voltage mediated dielectric tuning we are able to bring either two, or even three vibrational modes of the system in resonance. The resulting avoided crossings clearly demonstrate the underlying strong intermodal coupling. We employ an analytic model of four coupled harmonic oscillators to fit the measured avoided crossing diagrams and extract the observed coupling strengths. This model is also used to calculate the mode polarization, which reveals
in- and out-of-plane components of the hybridized normal modes at resonance.
\\
Figure \ref{1}a and b display scanning electron micrographs of the 180\,nm wide, 100\,nm thick and 53\,\textmu m long silicon nitride strings. The window structure in the shared clamping point is chosen such that the extension of the squared normalized local amplitude $\epsilon^2$ of the vibrational strain field induced in the clamping area is maximized. The window in the 16\,\textmu m x 16\,\textmu m large clamping point has dimensions of 4\,\textmu m x 8\,\textmu m of width and height. The central coupling beam and the window ledge are 200\,nm and 300\,nm in width, respectively. The device has an undercut of approx. 300\,nm, such that all parts of the window are freely suspended. The resonance frequencies of resonators 1 and 2 are found at $\omega_{1O}/2\pi$\,=\,6.37\,MHz, $\omega_{1I}/2\pi$\,=\,6.62\,MHz, $\omega_{2O}/2\pi$\,=\,6.50\,MHz and $\omega_{2I}/2\pi$\,=\,6.75\,MHz, where O corresponds to the out-of-plane mode and I to the in-plane mode, respectively. Even though both resonators are nominally identical such that their eigenfrequencies should coincide, we observe separate resonances for both resonators since fabricational imperfections result in a frequency mismatch of 1-2\,\%. The presence of the window does not affect the quality factor of the system which is about 200,000 in vacuum and at room temperature. Figure \ref{1}c shows finite element simulations of the squared normalized local amplitude $\epsilon^2$, which is indicative of the strain distribution in the clamping region. While the strain produced by the two resonators remains local without the window structure, the strain fields overlap in the window both for the IP and OOP mode allowing the resonators to exchange vibrational energy. The simulations also reveal that the presence of the window affects the magnitude of $\epsilon^2$, which is enhanced by two orders of magnitude for the OOP mode and by one order of magnitude for the IP mode.\\
We use standard interferometric displacement detection to simultaneously probe all four fundamental eigenfrequencies of the system. First we discuss the case of a two-mode resonance between the two resonators. Figure \ref{2} shows the frequency response of the resonator pair as a function of the DC voltage applied to the electrodes flanking resonator 1. The eigenmodes are dielectrically driven by RF voltage noise at -10\,dBm drive power, while the individual spectra are obtained by a spectrum analyzer. Variation of the DC voltage yields a quadratic dielectric tuning of two of the four frequency branches which allows to identify these branches with the eigenmodes of resonator 1. The softening/stiffening behavior allows to distinguish between IP and OOP polarization\cite{Rieger_Weig_2012}. Resonator 2 is actuated only indirectly via induced strain by the motion of resonator 1 since reference measurements have shown that for this resonator geometry the force of the stray electrical gradient field of the electrodes surrounding resonator 1 on resonator 2 is negligibly small. This is confirmed by numerical simulations which yield forces per unit length of $10^{-5}$\,N/m ($10^{-6}$\,N/m) for the OOP (IP) mode of resonator 1, but $10^{-8}$\,N/m for OOP and IP of resonator 2, what is at least two orders of magnitude smaller. A pronounced avoided crossing is apparent in Fig. \ref{2} at a DC voltage of about 17.2\,V, which can be assigned to the resonances between the IP mode of resonator 1 and the OOP mode of resonator 2 (1I and 2O, respectively). 
This identification is confirmed by an analytical model which extends the classical equations of motion of two strongly coupled, undamped harmonic oscillators\cite{Novotny} to the case of four oscillators with masses $m_i$, spring constants $k_i$ and complex coupling constants $\kappa_{ij}$ ($i,j=1,2,3,4$, where $i\neq j$ and $ij=ji$)
\begin{align}
\begin{split}
m_{i}\ddot{u}_{i}+k_{i}u_{i}+\sum_{i\neq j}\kappa_{ij}\left(u_{i}-u_{j}\right)=0
\end{split}
\label{eq_of_motion}
\end{align}
where $u_i$ is the deflection of the i-th oscillator. The common ansatz $u_{i}=u_{i}^{0}\exp\left(-i\omega t\right)$ is used
and we express $\kappa_{ij}$ in terms of the coupling strength\cite{Novotny} $g_{ij}$ with 
\begin{align}
g_{ij}=\frac{\kappa_{ij}}{\sqrt{\omega_{i}\omega_{j}m_{i} m_{j}}}.\label{eq_definition_Gamma}
\end{align}
For the case of the two nominally identical string resonators under investigation the effective masses of all four modes coincide, $m_i=0.5\rho V$ ($\rho V$ being the physical mass of each string), and we finally obtain a set of four equations for the angular eigenfrequencies $\omega_i$
\begin{align}
\omega_{i}^2&=\omega_{i0}^2+\sqrt{\omega_i}\left(\sum_{i\neq j}g_{ij}\sqrt{\omega_j}\right).
\label{eq:all_omega_i_expressed_with_Gamma}
\end{align}
Here $\omega_{i0}=\sqrt{k_i/m}$ are the tuning voltage ($U_{dc}$) dependent angular frequencies in absence of coupling $\omega_{i0}(U_{dc})=\omega^*_{i0}+2\pi(c_{i}(U_{dc}-U_{i0})^2+d_{i}(U_{dc}-U_{i0})^3)$
with $\omega^*_{i0}$ being the intrinsic angular eigenfrequency in absence of coupling and $c_{i}$ and $d_{i}$ are the quadratic tuning factor and a cubic correction. The indices $i$ and $j$ can be identified with the fundamental modes of the two resonators: $i$\,=\,1\,:\,1O, $i$\,=\,2\,:\,2O, $i$\,=\,3\,:\,1I and $i$\,=\,4\,:\,2I. Note that the dielectric tuning of the modes of resonator 1 is at least two orders of magnitude larger than for resonator 2 ($c_{1},c_{3}\gg c_{2},c_{4}$), as a consequence of the strongly suppressed dielectric force on resonator 2. The offset $U_{i0}$ denotes an eventual shift of the frequency tuning parabola, which can arise e.g. from a static polarization of a resonator.
\\
We employ a genetic algorithm\cite{Michaelewicz,Mitchell,Gerdes} to reliably fit this model to our data. To this end we extract $\omega_{i0}$ from our measured data and use $\omega^{*}_{i0}$, $c_{i}$,$d_{i}$, $U_{i0}$ and $g_{ij}$ as fit parameters.
Fitting our model (dashed blue lines in Fig. \ref{2}) to the measurement data we achieve very good agreement. We find a coupling strength $g_{2O,1I}$/2$\pi$\,=\,29.7\,kHz which exceeds the modes' linewidths in the range of 70-80\,Hz by more than two orders of magnitude, confirming that the system is deeply within the strong coupling regime.\\
We can also solve the eigenvalue problem of eq. \ref{eq:all_omega_i_expressed_with_Gamma} to obtain the corresponding eigenvectors $\vec{v}_i$
\begin{align}
\vec{u}_i(t) =\sum_{i\neq j} \vec{v}_i(t) \alpha \cos(\omega_j\cdot t + \phi_j),\label{eq:matrix}
\end{align}
(parameters $\alpha$ and $\phi_j$ are determined by initial conditions). The solution of the eigenvalue problem yields the contributions of resonator 1 and 2 to each polarization vector and thus to the mode polarization. When the modes are in resonance, we can distinguish between in-phase (both polarization vectors pointing along the positive, or negative direction of the I-O axes defined in Fig. \ref{2} ) and out-of-phase (one polarization vector pointing in negative and one in positive I-O axis direction) hybrid normal modes in analogy to the symmetric and antisymmetric normal modes of two coupled resonators\cite{Okamoto_Yamaguchi_2013}. In the following, we discuss the mode polarization of the four modes at the voltage of the avoided crossing, 17.2\,V, labelled A,B,C,D in Fig \ref{2}. As mode 2I does not contribute to the avoided crossing, it exhibits a clear in-plane mode polarization throughout the whole voltage sweep such that mode D can be directly identified with mode 2I. Similarly, mode 1O shows clear out-of-plane polarization and only starts to hybridize with mode 2O near 21\,V, at the edge of the accessible tuning range. Therefore mode A can be identified with mode 1O. Looking at the mode polarization (black arrows) in the avoided crossing, we find that mode C belongs to a hybrid in-phase mode and mode B belongs to a hybrid out-of-phase mode. So clearly the observed avoided crossing is reminiscent of strong inter-string coupling, confirming that the engineered clamping region enhances the strain-mediated intrinsic coupling of the two strings.
\begin{figure}[t]
\includegraphics{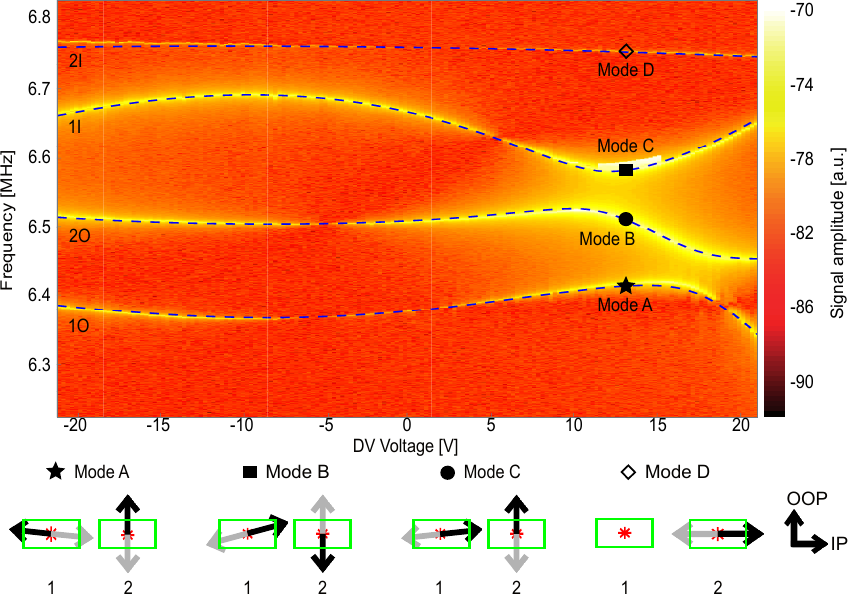}
\caption{\label{3}(color online) Three-mode resonance: DC voltage sweep of resonator 1 shows multimode avoided crossings between 1O, 2O and 1I, giving rise to the hybrid modes A, B and C. Mode 2I is not involved in the 3-mode resonance (mode D). Signal amplitude is plotted on a logarithmic color scale. The analytic model (blue dashed lines) yields coupling strengths of $g_{1O,2O}/2\pi$=\,130\,kHz and $g_{2O,1I}/2\pi$=\,60\,kHz, as well as $g_{1O,1I}/2\pi$=\,50\,kHz. The black arrows show the respective mode polarization ($\updownarrow$ : OOP, $\leftrightarrow$ : IP) for a DC voltage of 13.4\,V, grey arrows show the inverse.}
\end{figure}
\\
Depending on the intrinsic frequency differences between the modes and the quadratic tuning behavior of the resonances it is possible to get one step further and demonstrate a 3-mode resonance of the two resonators. Figure \ref{3} shows a voltage sweep from -21 to 21\,V. Note that the spectra in Fig. \ref{2} and Fig. \ref{3} have been obtained from the same resonator pair. The transition from simple 2-mode to 3-mode resonance is enabled by a shift of the tuning curves of resonator 1, i.e. an offset of the vertices $U_{10}$ and $U_{30}$ induced by exposing the resonator to a large DC voltage over 24 hours. 
This shift is attributed to the buildup of static dipoles in the SiN material of resonator 1 and can be reversed by venting the vacuum chamber.\\
The measurement shown in Fig. \ref{3} was performed with a vector network analyzer at a constant drive power of -10\,dBm. We find avoided crossings between mode 2O and 1I at about 11\,V, between mode 1O and 1I between 11 and 16\,V, as well as an avoided crossing between modes 1O and 2O at about 16\, V. Note that the coupling strengths in the 3-mode resonance can not be easily set equivalent to the frequency splitting between the modes. Fitting eq. \ref{eq:all_omega_i_expressed_with_Gamma} to the data yields coupling strengths $g_{1O,2O}/2\pi$=\,130\,kHz, $g_{2O,1I}/2\pi$=\,60\,kHz and $g_{1O,1I}/2\pi$=\,50\,kHz, again all deep in the strong coupling regime. Mode 2I remains far from resonance with the other modes. We also observe, that the resonances become highly nonlinear in the coupling region as apparent from their asymmetric intensity profiles.\\ 
Since the vibrational modes of resonator 2 are only very slightly affected by the electrical gradient field, we can assume the coupling between modes from resonator 1 and 2 to be of pure mechanical character, whereas the intermodal coupling of resonator 1 is dielectrical\cite{Faust_2012}. Hence the 3-mode resonance reveals that both coupling channels can be combined to generate multi-mode coupling architectures.\\
Looking at the calculated mode polarizations at 13.4\,V (Fig. \ref{3}), i.e. in the region of the 3-mode resonance, we find out-of-phase hybrid modes A and B, with opposing relative phase. Mode C can be identified as an in-phase hybrid mode. Overall considerations of the mode polarizations throughout the whole voltage sweep, which are not shown here, indicate that the lowest branch starts out as a pure out-of-plane mode (1O) at -21\,V, hybridizes into mode A and ends up an in-plane mode (1I) at 21\,V. The third branch shows the reverse behavior: starting from in-plane (1I) at -21\,V, hybridizing into mode C and turning into an out-of-plane mode (1O) when reaching 21\,V. For the second branch the transition through the coupling region is more complex: starting out as an out-of-plane mode (2O) at -21\,V it starts to hybridize with 1I before transforming into mode B, which in turn hybridizes with 1O and finally transforms back into an out-of-plane mode (2O). Note that the fourth branch remains the 2I mode throughout the whole sweep, since it can not be tuned in resonance with any other mode, such that mode D can be identified with 2I.
\begin{figure}[t]
\includegraphics{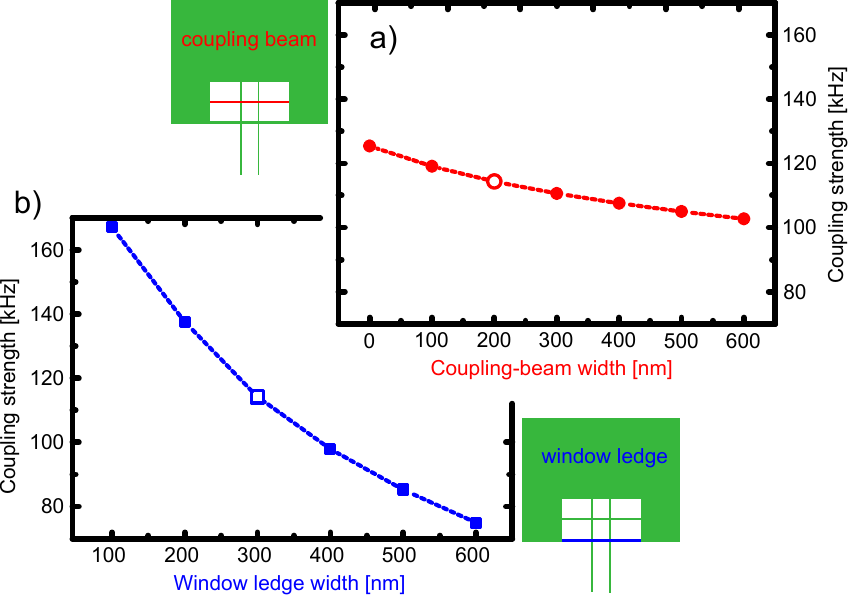}
\caption{\label{4}(color online). a: Finite element simulation of the OOP-OOP coupling strength for different coupling beam widths from 0 to 600\,nm. b: Simulated coupling strength for window ledge widths from 100 to 600\,nm. For the window geometry under investigation (coupling beam width 200\,nm and window ledge 300\,nm) a coupling strength of $g_{1O,2O}/2\pi$=\,115\,kHz is obtained (hollow markers), which agrees well with the experimentally observed $g_{1O,2O}/2\pi$=\,130\,kHz (Fig. \ref{3}).}
\end{figure}\\
In order to further tune and manipulate the resonator system, it is necessary to control not only the dielectric intra-resonator but also the strain-mediated inter-resonator coupling strengths. Therefore we numerically explore the impact of the window geometry on the out-of-plane and in-plane frequencies of two identical resonators. Figure \ref{4} shows a finite element simulation of the coupling strength between the out-of-plane modes of the resonators as a function of the coupling beam width (Fig. \ref{4}a) and the lower window ledge width (Fig. \ref{4}b). The coupling of these modes corresponds to the coupling of modes 1O and 2O in our system. The geometry and dimensions of the simulated structure match the experimentally explored device for the case of the hollow data point in Fig. \ref{4}a and b. The widths of the coupling beam and the window ledge are varied from 100 to 600\,nm and we also consider the case of no coupling beam (0\,nm). Note that a coupling beam or window ledge wider than 600\,nm will no longer be freely suspended for the given undercut of 300\,nm, which will immediately decrease the coupling to near zero (not shown). For decreasing window ledge width we find a strong increase of the coupling strength from 80\,kHz to 160\,kHz in Fig. \ref{4}b. For decreasing coupling beam width, the coupling strength is predicted to increase from 100\,kHz to 130\,kHz according to Fig. \ref{4}a. For the experimental sample dimensions of a window ledge width of 300\,nm and a coupling beam width of 200\,nm, the simulated coupling strength is $g_{1O,2O}/2\pi$=\,115\,kHz. This agrees very well with the experimentally obtained $g_{1O,2O}/2\pi$=\,130\,kHz (see Fig. \ref{3}). 
Hence, changing the coupling beam and window ledge widths will allow to tune the inter-resonator coupling strength. In combination with a careful choice of the resonator dimensions as well as the electrical field gradients, this will allow to adjust all parameters of the multi-mode system.\\
In conclusion, we demonstrated strong strain-mediated intrinsic coupling between two silicon-nitride string resonators via an engineered joint clamping point. Tuning the resonance frequencies of the first, dielectrically driven resonator allows to bring two or even three modes of the two resonators in resonance, showing multiple avoided crossings. An analytical model describes the behavior of the coupled system with excellent agreement and enables us to evaluate not only the coupling strength, but also the polarization of the vibrational normal modes even in complex coupling situations. Simulations show that the inter-resonator coupling strength of the system can be controlled by varying the widths of the coupling beam and the window ledge. Future experiments will include a third electrode flanking resonator 2, which will allow for independent frequency tuning of both resonators. Ultimately, the resulting architecture will allow to adjust the coupling behavior of arbitrary modes at will, enabling the realization of larger nanomechanical arrays\cite{Buks_Roukes_2002,Huang_Du_2016} for studying the dynamics of resonator networks. Such resonator networks are of immediate interest not only for the study of synchronization effects, but also for exploring the coherent dynamics of mechanical multimode systems, including mode propagation or frequency and mode transformation. In particular, tailoring the resonator interaction as well as the array geometry will allow to realize mechanical metamaterials with the potential to explore non-reciprocal properties\cite{Huang_Du_2016} or topological effects\cite{Salerno_2017,Huber_2015,Peano_Marquardt_2015}. 
\begin{acknowledgments}
We like to thank Fabian M\"uller for his help setting up the genetic fitting algorithm and Benjamin Gmeiner as well as Onur Basarir for preliminary experiments.
This project has received funding from the European
Union's Horizon 2020 research and innovation program
under grant agreement No 732894.
Financial support by the Deutsche Forschungsgemeinschaft via the collaborative research center SFB\,767 is gratefully acknowledged. 
\end{acknowledgments}

% If in two-column mode, this environment will change to single-column format so that long equations can be displayed. 
%\begin{widetext}
%$$\mbox{put long equation here}$$
%\end{widetext}

% Figures should be put into the text as floats. 
% Use the graphics or graphicx packages (distributed with LaTeX2e).
% Put the label that you will use with \ref{} command in the braces of the \label{} command.
%
% \begin{figure}
% \end{figure}

% \begin{table}
% \end{tabular}
% \end{table}

%\begin{acknowledgments}
% Put your acknowledgments here.
%\end{acknowledgments}

%\bibliography{coupled-resonators}

%\bibliography{coupled-resonators}

%

\end{document}